\documentclass[aps,pra,twocolumn,showpacs]{revtex4}
\usepackage{graphicx}
\usepackage{dcolumn}
\usepackage{bm}
\usepackage{amsfonts}
\usepackage{amsmath}

\begin{document}
\renewcommand{\thefigure}{\arabic{figure}}

\title{Emergence of Wigner molecules in one-dimensional systems of repulsive fermions under harmonic confinement}
\author{Saeed H. Abedinpour}
\author{Marco Polini}
\email{m.polini@sns.it}
\author{Gao Xianlong}
\author{M. P. Tosi}
\affiliation{NEST-CNR-INFM and Scuola Normale Superiore, I-56126 Pisa, Italy}
\date{\today}

\begin{abstract}
A Bethe-{\it Ansatz} spin-density functional approach is developed to evaluate the ground-state density profile in a system of repulsively interacting spin-$1/2$ fermions inside a quasi-one-dimensional harmonic well. The approach allows for the formation of antiferromagnetic quasi-order with increasing coupling strength and reproduces with high accuracy the exact solution that is available for the two-fermion system.
\end{abstract}
\maketitle

{\it Introduction} ---It is known from bosonization~\cite{ref:giamarchi_book} that the ground-state $|{\rm GS}\rangle$ of a one-dimensional ($1D$) system of repulsively interacting spin-$1/2$ fermions carries antiferromagnetic quasi-ordering. Quantum fluctuations prevent any type of true spin-symmetry breaking in $1D$ and $|{\rm GS}\rangle$ may be viewed as a superposition of two spin-density waves (SDW),
\begin{equation}\label{eq:sdw}
|{\rm GS}\rangle \propto |\uparrow\downarrow\uparrow\downarrow...\rangle+ |\downarrow\uparrow\downarrow\uparrow...\rangle
\end{equation}
so that the local spin polarization is rigorously zero. Nevertheless, ordering is signaled by the co-called ``$2k_F\rightarrow 4k_F$ crossover" in the wave number $q$ of bulk Friedel oscillations: for vanishing and weak interactions the ground state is liquid-like with $q=2k_{\rm F}$, whereas at strong repulsions a periodicity $q=4 k_{\rm F}$ emerges. Here $k_{\rm F}$ is the Fermi wave number, related to the fermion density $n$ by $k_{\rm F}=\pi n/2$, and each modulation is weighted by a slow decay factor. 

In previous work~\cite{ref:gao_pra_2006} we studied a system of $N$ spin-$1/2$ fermions subject to a 
strongly anisotropic harmonic potential, characterized by angular frequencies $\omega_\perp$ and $\omega_\|$ in the radial and axial directions with $\omega_\| \ll \omega_\perp$. The fermions are dynamically $1D$ if the anisotropy parameter of the confinement is much smaller than the inverse particle number, $\omega_{\|}/\omega_\perp\ll N^{-1}$.
The two species of fermionic particles are taken to have the same mass $m$, to be in equal numbers $N_\uparrow$ and $N_\downarrow$, and to interact {\it via} a contact repulsion with an effective $1D$ coupling strength $g_{\rm 1D}\geq 0$~~\cite{ref:olshanii_1998}. The system is described by the Hamiltonian
\begin{equation}\label{eq:igy}
{\cal H}=-\frac{\hbar^2}{2m}\sum_{i=1}^{N}\frac{\partial^2}{\partial z^2_i}+g_{\rm 1D}\sum_{i=1}^{N_{\uparrow}}\sum_{j=1}^{N_{\downarrow}}\delta(z_i-z_j)+V_{\rm ext}\,,
\end{equation}
where contact interactions between parallel-spin particles are suppressed by the Pauli principle and 
$V_{\rm ext}=\sum_{i=1}^{N}
V_{\rm ext}(z_i)=(m\omega^2_{\|}/2)\sum_{i=1}^{N}z^2_i$ is the external static potential associated 
with the axial confinement. Choosing the harmonic-oscillator length $a_{\|}=\sqrt{\hbar/(m\omega_{\|})}$ as unit of length and the harmonic-oscillator quantum $\hbar\omega_{\|}$ as unit of energy, 
the Hamiltonian (\ref{eq:igy}) can be shown to be governed by 
the dimensionless coupling parameter 
\begin{equation}\label{eq:lambda}
\lambda=\frac{g_{\rm 1D}}{a_{\|}\hbar\omega_{\|}}\,.
\end{equation}
For $V_{\rm ext}=0$ the Hamiltonian (\ref{eq:igy}) reduces to the homogeneous Gaudin-Yang model, which is exactly solvable by means of the Bethe-{\it Ansatz} technique~\cite{ref:GY}.

In Ref.~\onlinecite{ref:gao_pra_2006} a density-functional scheme using as reference the Bethe-{\it Ansatz} homogeneous fluid was proposed to calculate the ground-state properties of the inhomogeneous $1D$ system described by the Hamiltonian (\ref{eq:igy}). However, this approach was found to fail at strong-coupling (see Fig.~8 at $\lambda=10$ in Ref.~\onlinecite{ref:gao_pra_2006}): more precisely, the exchange-correlation potential proposed there is unable to describe the $2k_{\rm F}\rightarrow 4k_{\rm F}$ crossover, which as noted above is expected to occur upon increasing the strength of the repulsive interactions between antiparallel-spin particles. In the present note we propose a simple functional that embodies this crossover and is capable of describing inhomogeneous Luttinger liquids with strong repulsions. The main idea is to capture the tendency to antiferromagnetic spin ordering (i) by adding an infinitesimal spin-symmetry-breaking magnetic field to the Hamiltonian (\ref{eq:igy}),
\begin{equation}\label{eq:mfield}
V_{\rm ext}\rightarrow \frac{1}{2}m\omega^2_{\|}\sum_{i=1}^N z^2_i-\mu_B\sum_{i=1}^N \sigma^z_i B(z_i)\,,
\end{equation}
with $\mu_B$ the Bohr magneton; and (ii) by resorting to spin-density functional theory (SDFT)~\cite{ref:Giuliani_and_Vignale}.

{\it The ground-state energy of the Gaudin-Yang model for finite $\langle {\hat S}_z\rangle$} ---The addition of a magnetic field to the Hamiltonian (\ref{eq:igy}) as prescribed by Eq.~(\ref{eq:mfield}) requires that we know the ground-state energy of the homogeneous Gaudin-Yang model for the situation $N_\uparrow \neq N_\downarrow$, which will be our reference system with Luttinger-liquid ground-state correlations. In the thermodynamic limit ($N_\uparrow,N_\downarrow$, and $L \rightarrow \infty$, $L$ being the system size) the properties of the reference fluid are determined by the linear density $n=N/L$, by the spin polarization $\zeta=(N_\uparrow-N_\downarrow)/N$, and by the effective coupling $g_{\rm 1D}$. The linear density and the coupling parameter can be conveniently combined into a single dimensionless parameter $\gamma \equiv mg_{\rm 1D}/(\hbar^2 n)$. 

The energy per particle can be written in terms of the momentum distribution $\rho(k)$ as
\begin{equation}\label{eq:energy_atom}
\varepsilon_{\rm GS}(n,\zeta,g_{\rm 1D})=
\frac{1}{n}\int_{-Q}^{+Q}dk\,\frac{\hbar^2 k^2}{2m}\rho(k)\,.
\end{equation}
The function $\rho(k)$ can be calculated by solving the two-coupled Gaudin-Yang Bethe-{\it Ansatz} integral equations~\cite{ref:GY},
\begin{equation}\label{eq:rho}
\rho(k)=\frac{1}{2\pi}+\frac{2\gamma n}{\pi}\int_{-B}^{B}dq\,\frac{\sigma(q)}{(n\gamma)^2+4(k-q)^2}
\end{equation}
and
\begin{eqnarray}\label{eq:sigma}
\sigma(k)&=&-\frac{\gamma n}{\pi}\int_{-B}^{B}dq\,\frac{\sigma(q)}{(n \gamma)^2+(k-q)^2}\nonumber\\
&+&\frac{2\gamma n}{\pi}\int_{-Q}^{Q}dq\,\frac{\rho(q)}{(n \gamma)^2+4(k-q)^2}\,,
\end{eqnarray}
where $Q$ and $B$ are determined by the normalization conditions
\begin{equation}\label{eq:normalization_1}
\int_{-Q}^{+Q}\rho(k)dk=n
\end{equation}
and
\begin{equation}\label{eq:normalization_2}
\int_{-B}^{+B}\sigma(k)dk=\frac{n}{2}(1-\zeta)\,.
\end{equation}
We have fitted the results for $\varepsilon_{\rm GS}(n,\zeta,g_{\rm 1D})$, obtained from the numerical solution of the 
Gaudin-Yang equations (\ref{eq:rho})-(\ref{eq:normalization_2}), with a simple yet accurate parametrization formula given below.

\begin{figure}
\begin{center}
\includegraphics[width=1.00\linewidth]{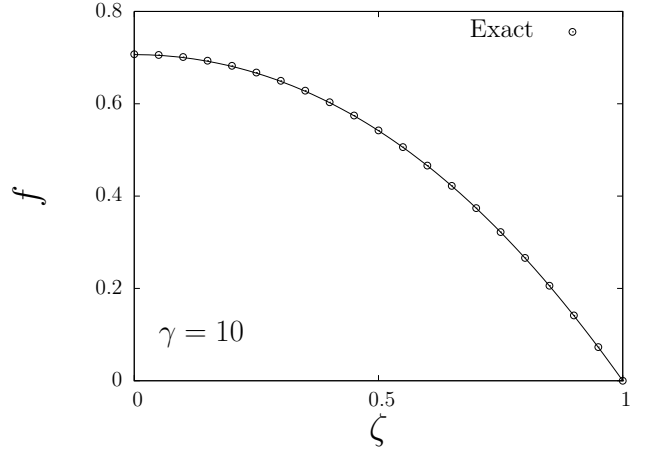}
\caption{The interaction contribution $f$ to the ground-state energy 
$\delta \varepsilon_{\rm GS}(n,g_{\rm \scriptscriptstyle 1D})$ 
of the homogeneous Gaudin-Yang model (see text) as a function of the spin polarization $\zeta$ for $\gamma=10$. The exact result, obtained from the solution of the Bethe-Ansatz equations (\ref{eq:rho})-(\ref{eq:normalization_2}), is compared with the fitting formula in Eq.~(\ref{eq:fit}).\label{fig:one}}
\end{center}
\end{figure}

We introduce the Fermi wave number $k_{\rm F}=\pi n/2$ of the unpolarized system and the corresponding Fermi energy $\varepsilon_{\rm F}=\hbar^2 k^2_{\rm F}/(2m)$. We write
\begin{equation}
\varepsilon_{\rm GS}(n,\zeta,g_{\rm 1D})=\kappa(n,\zeta)+\delta \varepsilon_{\rm GS}(n,\zeta,g_{\rm 1D})
\end{equation}
where
\begin{equation}
\kappa(n,\zeta)=\frac{\pi^2\hbar^2n^2}{24m}(1+3\zeta^2)
\end{equation}
is the kinetic energy of the noninteracting system per particle. We find that $\delta \varepsilon_{\rm GS}(n,\zeta,g_{\rm 1D})$ in units of the Fermi energy, $f \equiv \delta \varepsilon_{\rm GS}(n,\zeta,g_{\rm 1D})/\varepsilon_{\rm F}$, can be very accurately represented by the parametrization
\begin{eqnarray}\label{eq:fit}
f(x,\zeta)&=&[e(x)-1/3]\left\{1+\alpha(x)\zeta^2+\beta(x)\zeta^4\right.\nonumber\\
&-&\left.[1+\alpha(x)+\beta(x)]\zeta^6\right\}
\end{eqnarray}
where $x=2\gamma/\pi$, $e(x)$ is given in Eq.~(21) of Ref.~\onlinecite{ref:gao_pra_2006}, and
\begin{equation}
\left\{
\begin{array}{l}
{\displaystyle \alpha(x)=\frac{-x^2+a_\alpha x+b_\alpha}{x^2+c_\alpha x-b_\alpha}}
\vspace{0.1 cm}\\
{\displaystyle \beta(x)=\frac{a_\beta x}{x^2+b_\beta x+c_\beta}}
\end{array}
\right.\,.
\end{equation}
Here $a_\alpha=-1.68894$, $b_\alpha=-8.0155$, $c_\alpha= 2.74347$, $a_\beta=-1.51457$, $b_\beta=2.59864$, and $c_\beta=6.58046$. Equation~(\ref{eq:fit}) is compared with the exact Bethe-Ansatz results for $\gamma=10$ in Fig.~\ref{fig:one}.

{\it Spin-density-functional theory in the Kohn-Sham scheme} ---The ground-state spin-densities $n_\sigma(z)$ 
can be calculated within the Kohn-Sham version of SDFT by solving self-consistently the effective Schr\"odinger equations for single-particle orbitals
\begin{equation}\label{eq:kss}
\left[-\frac{\hbar^2}{2m}\frac{\partial^2}{\partial z^2}+V^{(\sigma)}_{\rm KS}[n_\sigma](z)\right]\varphi_{\alpha,\sigma}(z)=\varepsilon_{\alpha,\sigma}\varphi_{\alpha,\sigma}(z)
\end{equation}
with $V^{(\sigma)}_{\rm KS}[n_\sigma](z)=V^{(\sigma)}_{\rm H}[n_\sigma](z)+V^{(\sigma)}_{\rm xc}[n_\sigma](z)+V^{(\sigma)}_{\rm ext}(z)$, together with the closure
\begin{equation}\label{eq:closure}
n_\sigma(z)=\sum_{\alpha, {\rm occ.}}\Gamma^{(\sigma)}_\alpha\left|\varphi_{\alpha,\sigma}(z)\right|^2\,.
\end{equation}
Here the sum runs over the occupied orbitals and the degeneracy factors $\Gamma^{(\sigma)}_\alpha$ satisfy the sum rule 
$\sum_\alpha \Gamma^{(\sigma)}_\alpha=N_\sigma$. The first term in the spin-dependent effective Kohn-Sham potential $V^{(\sigma)}_{\rm KS}$ is the mean-field term $V^{(\sigma)}_{\rm H}=g_{\rm 1D}n_{\bar \sigma}(z)$, while the second term is the exchange-correlation potential defined as the functional derivative of the exchange-correlation energy $E_{\rm xc}[n_\sigma]$ evaluated at the ground-state density profile, $V^{(\sigma)}_{\rm xc}=\delta E_{\rm xc}[n_\sigma]/\delta n_\sigma(z)|_{\rm \scriptscriptstyle GS}$. Finally, the third term is the spin-dependent external field $V^{(\sigma)}_{\rm ext}(z)=m\omega^2_{\|}z^2/2-\mu_B\sigma B(z)$.

Equations~(\ref{eq:kss}) and~(\ref{eq:closure}) provide a formally exact scheme to calculate $n_\sigma(z)$, but $E_{\rm xc}$ needs to be approximated. The local-spin-density approximation (LSDA) is known to provide an excellent description of the ground-state properties of a variety of inhomogeneous systems~\cite{ref:Giuliani_and_Vignale}. In the following we employ a Bethe-Ansatz-based LSDA functional (BALSDA) for the exchange-correlation potential,
\begin{eqnarray}\label{eq:balda}
E_{\rm xc}[n_\sigma] & \rightarrow & E^{\rm BALSDA}_{\rm xc}[n_\sigma]\\
&=&\int dz\, n(z)\left.\varepsilon^{\rm hom}_{\rm xc}(n,\zeta,g_{\rm 1D})\right|_{n\rightarrow n(z),\zeta\rightarrow \zeta(z)}\nonumber
\end{eqnarray}
where the exchange-correlation energy $\varepsilon^{\rm hom}_{\rm xc}$ of the homogeneous Gaudin-Yang model is defined by
\begin{eqnarray}\label{eq:xc}
\varepsilon^{\rm hom}_{\rm xc}(n,\zeta,g_{\rm 1D})&=&\varepsilon_{\rm GS}(n,\zeta,g_{\rm 1D})-\kappa(n,\zeta)\nonumber\\
&-&\varepsilon_{\rm H}(n,\zeta,g_{\rm 1D})\,.
\end{eqnarray}
Here 
\begin{equation}
\varepsilon_{\rm H}(n,\zeta,g_{\rm 1D})=\frac{1}{4}g_{\rm 1D}n^2(1-\zeta^2)
\end{equation}
is the mean-field energy.

{\it Illustrative numerical results for the two-body problem} ---To illustrate how this procedure based on SDFT and BALSDA embodies the $2k_F\rightarrow 4 k_F$ crossover we present some numerical results for the case $N=2$, which is exactly solvable as shown in Ref.~\onlinecite{ref:busch_huyel_2003} and in the Appendix of Ref.~\onlinecite{ref:gao_pra_2006}.

For the case $N=2$ we choose the following form for the infinitesimal spin-symmetry-breaking magnetic field:
\begin{equation}
B(z)=\frac{2\epsilon}{\pi}\arctan{(z/\xi)}\,,
\end{equation}
where the field strength $\epsilon$ is reduced to zero during the self-consistent solution of the Kohn-Sham equations and $\xi$ is a parameter that we take as $\xi=0.5 a_{\|}$. Two important remarks can be made: (i) if $\lambda$ is small the addition of this infinitesimal field is irrelevant and, at the end of the self-consistency cycle, 
one reproduces the results presented in the top and central panels of Fig.~8 in Ref.~\onlinecite{ref:gao_pra_2006}; and (ii) if $\lambda$ is large instead, the final result for the total density $n(z)$ is very different from that resulting from the BALDA theory of Ref.~\onlinecite{ref:gao_pra_2006}. We have carefully checked that the final result is independent of the special form of the spin-symmetry-breaking field. In fact one could alternatively start the self-consistent Kohn-Sham cycle from an initial guess which has slightly broken spin sysmmetry ({\it i.e.} from $n_\uparrow(z)\neq n_\downarrow(z)$), without using any $B$ field.

\begin{figure}
\begin{center}
\includegraphics[width=1.0\linewidth]{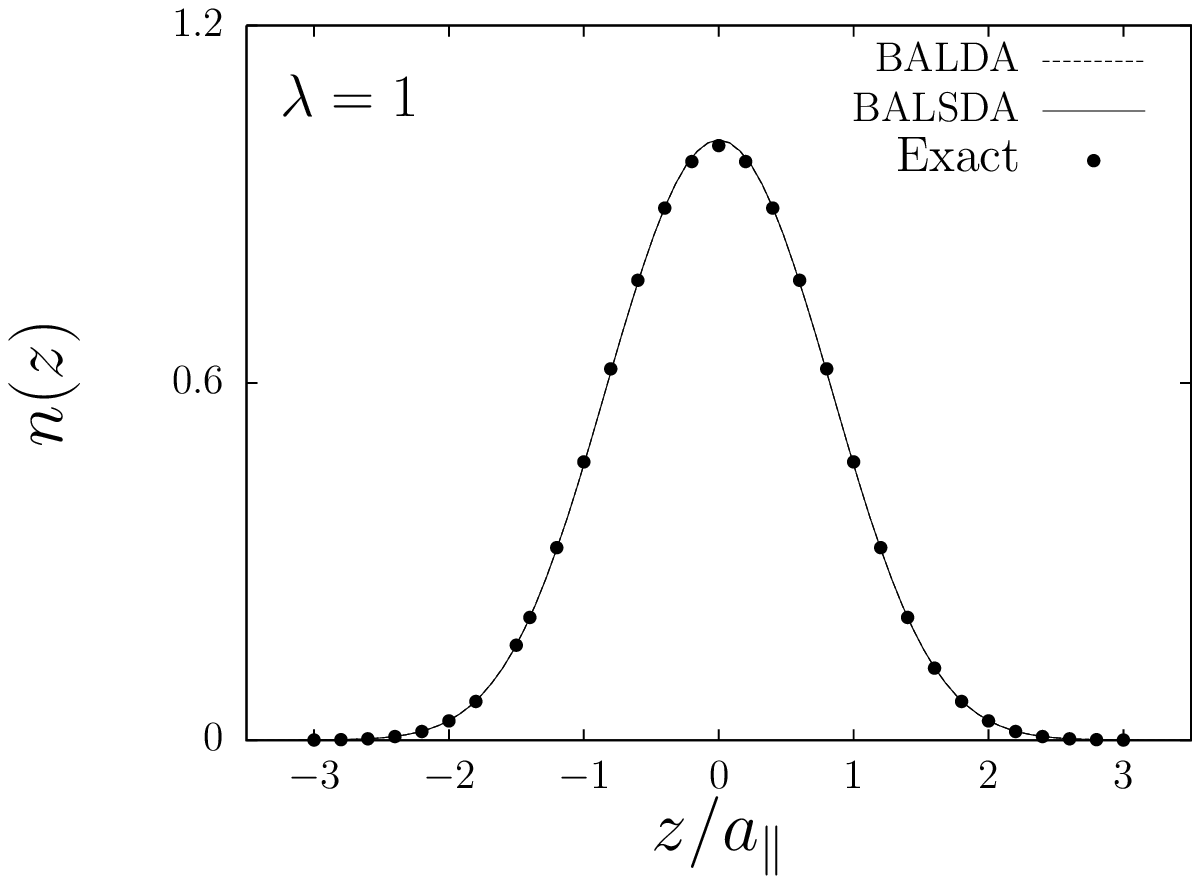}\\
\includegraphics[width=1.0\linewidth]{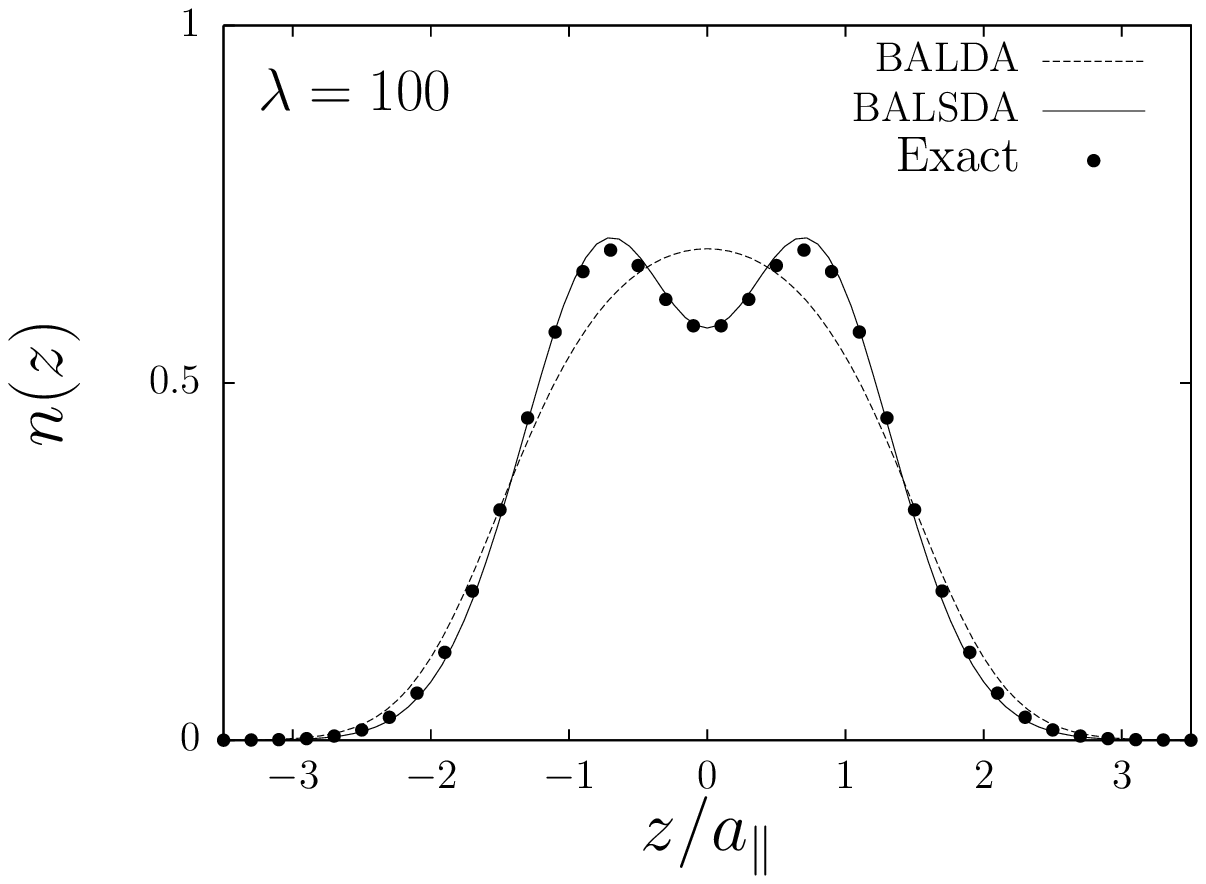}
\caption{Density profile $n(z)$ (in units of $a_{\|}^{-1}$) as a function of $z/a_{\|}$ 
for two fermions with antiparallel spins at $\lambda=1$ (top) and $\lambda=100$ (bottom). 
The results of the BALSDA scheme proposed in this work are compared with the exact solution of the two-body problem 
and with the BALDA results of Ref.~\onlinecite{ref:gao_pra_2006}. At $\lambda=1$ the BALSDA result coincides with the BALDA one.\label{fig:two}}
\end{center}
\end{figure}

Figure~\ref{fig:two} reports our numerical results for the total density profile of $N=2$ fermions with antiparallel spins. The emergence of a Wigner molecule at strong coupling is evident and very well described by the BALSDA approach.

{\it Acknowledgments} --- We gratefully acknowledge R. Asgari, K. Capelle, P. Capuzzi, and G. Vignale for several useful discussions.


\begin{thebibliography}{99}
\bibitem{ref:giamarchi_book}
	T. Giamarchi, {\it Quantum Physics in One Dimension} (Clarendon Press, Oxford, 2004).
\bibitem{ref:gao_pra_2006}
	Gao Xianlong, M. Polini, R. Asgari, and M.P. Tosi, \pra {\bf 73}, 033609 (2006).
\bibitem{ref:olshanii_1998}
	M. Olshanii, \prl {\bf 81}, 938 (1998); 
	T. Bergeman, M.G. Moore, and M. Olshanii, {\it ibid.} {\bf 91}, 163201 (2003).	
\bibitem{ref:GY}
	M. Gaudin, Phys. Lett. {\bf 24A}, 55 (1967); C.N. Yang, \prl {\bf 19}, 1312 (1967).
\bibitem{ref:Giuliani_and_Vignale}
	G.F. Giuliani and G. Vignale, {\it Quantum Theory of the Electron Liquid} 
	(Cambridge University Press, Cambridge, 2005).
\bibitem{ref:busch_huyel_2003}
	Th. Busch and C. Huyel, J. Phys. B {\bf 36}, 2553 (2003).
\end{thebibliography}
\end{document}